%% file: main.tex
\documentclass[trackchanges,twocolumn]{aastex63} %linenumbers

\pdfoutput=1 %for arXiv submission
\usepackage{amsmath,amstext,amssymb}
\usepackage{float}
\usepackage[T1]{fontenc}
\usepackage{xcolor}
\usepackage{soul}
\usepackage{ulem}
\usepackage{comment}
\usepackage{graphicx}
\setstcolor{red}
\usepackage{lineno}
\usepackage{appendix}
\usepackage[artemisia]{textgreek}
\usepackage{lipsum}
\usepackage{lineno}
\usepackage{cleveref}
\usepackage{mathtools}
\usepackage{enumitem}
\turnoffedit

\DeclareMathOperator{\NormalDist}{N}

\DeclareMathOperator{\PoiDist}{Poisson}

\undef{\listofchanges}
\undef{\todo}
\usepackage[commandnameprefix=always,commentmarkup=uwave]{changes} 
\definechangesauthor[name={WF}, color=cyan]{WF}

\graphicspath{{figures/}}

\newcommand{\appropto}{\mathrel{\vcenter{
  \offinterlineskip\halign{\hfil$##$\cr
    \propto\cr\noalign{\kern2pt}\sim\cr\noalign{\kern-2pt}}}}}

\shortauthors{Roy et al.}

\received{\today}

\begin{document}
\title{Cosmology with Binary Neutron Stars: Does Mass-Redshift Correlation Matter?}

\shorttitle{Cosmology with BNS: Does Mass-Redshift Correlation Matter?}

%\linenumbers
\input{authors}

\correspondingauthor{Soumendra Kishore Roy}
\email{soumendrakisho.roy@stonybrook.edu}

\begin{abstract}
    
   Next-generation gravitational wave detectors are expected to detect millions
   of compact binary mergers across cosmological distances. The features of
   the mass distribution of these mergers, combined with
   gravitational wave distance measurements, will
   enable precise cosmological inferences, even without the need for
   electromagnetic counterparts. However, achieving
   accurate results requires
   modeling the mass spectrum, particularly
   considering possible redshift evolution. Binary neutron star (BNS) mergers are thought to be less influenced by changes in metallicity compared to binary black holes (BBH) or neutron star-black hole (NSBH) mergers. This stability in their mass spectrum over cosmic time reduces the chances of introducing biases in cosmological parameters caused by redshift evolution. In this study, we use the population synthesis code
   \texttt{COMPAS} to generate astrophysically motivated catalogs of BNS mergers
   and explore whether assuming a non-evolving BNS mass distribution
   with redshift could introduce biases in cosmological
   parameter inference. \edit1{Our findings show that despite significant variations in the BNS mass distribution across binary physics assumptions and initial conditions in \texttt{COMPAS}, the joint mass-redshift population can be expressed as the product of the mass distribution marginalized over redshift and the redshift distribution marginalized over masses. This enables a $2\%$ unbiased constraint on the Hubble constant—sufficient to address the Hubble tension}. Additionally, we show that in the fiducial \texttt{COMPAS} setup, the bias from a non-evolving BNS mass model is less than $0.5\%$ for the Hubble parameter measured at redshift $0.4$. These results establish BNS mergers as strong
   candidates for spectral siren cosmology in the era of next-generation
   gravitational wave detectors. 
    
\end{abstract}


%%%%%%%%%%%%%%%%%%%%%%%%%%%%%%%%%%%%%%%%%%%%%%%%%%%%%%%%
\section{Introduction}\label{sec:intro}

Gravitational waves (GW) from compact binary coalescences (CBC) can measure the expansion rate of the Universe, denoted as $H(z)$ (where $z$ is redshift) \edit1{\citep{1986Natur.323..310S}}. This has gained significant interest in the scientific community, particularly due to its potential to address the $\sim 5 \sigma$ tension in the Hubble constant, $H_0$ \citep{Planck18, Riess_2022}, without relying on the cosmic distance ladder \citep{1986Natur.323..310S, 2005ApJ...629...15H, Farr_2019}. Currently, $H_0$ is constrained to approximately $20\%$ based on $47$ confident detections of CBCs after the third gravitational wave transient (GWTC-3) catalog \citep{Abbott_2023} by the LIGO-Virgo-KAGRA (LVK) collaboration \citep{aligo, avirgo, 10.1093/ptep/ptaa125}. The current measurements are dominated by the bright siren GW170817 \citep{Abbott_2017}, with the coincident detection of its electromagnetic counterpart significantly improving sky localization and facilitating the estimation of the redshift of the host galaxy. However, the probability of coincident detection is low and further restricts us to a smaller sky volume, even in the era of next generation (xG) GW detectors. Consequently, in the absence of an electromagnetic counterpart or galaxy catalog, alternative techniques become necessary to estimate redshift solely from GWs.

Estimating redshift only from GWs is \edit1{hard and} one of the most promising techniques relies on the fact that GWs redshift mass scales (for an alternative approach using neutron star equation of state to infer redshift, see \citet{Messenger_2012}). Identifying features, such as peaks or dips, in the mass distribution at different luminosity distances can estimate redshifts from a catalog of CBCs \citep{Chernoff_1993, Taylor_2012, Taylor_2012_1, Farr_2019}, a method known as the spectral-siren method \citep{Ezquiaga_2022}:
\begin{equation}\label{eq:redshift}
    z = \frac{m^{\text{det}}}{m}-1
\end{equation}
Here, `$m$' represents the source-frame mass in proximity to the feature, while `$m^{\text{det}}$' indicates the detected mass obtained using GWs near the said feature. This approach assumes that the mass distribution features do not change with redshift \citep{Farr_2019, PhysRevD.104.062009}. However, this assumption is likely to be violated at some level, as features in the mass spectra of CBCs are influenced by various astrophysical conditions that vary with redshifts \citep[e.g.,][]{2019MNRAS.487....2M, redshift_ev3, 2022MNRAS.511.5797M, Mukherjee_2022, Karathanasis_2023,2024ApJ...967...62Y}. 
Such variation can introduce bias in the redshift measurement, leading to an erroneous inference of $H(z)$ with a sufficient number of observations \citep{Ezquiaga_2022, pierra2023studysystematicscosmologicalinference}.

\edit1{In this work, we investigate whether the assumption of redshift independence of the BNS mass spectrum introduces bias in the inference of cosmological parameters.} It is, at present, impossible to observationally assess the redshift evolution of the BNS population because GWTC-3 contains only the two BNS mergers GW170817 and GW190425 \citep{gwtc-3pop}.  The limited sensitivity, and associated small search volumes, for the present generation of detectors will make measurements of redshift evolution difficult even with larger catalogs.

Next-generation (xG) observatories are expected to overcome this limitation, as their higher sensitivity will produce vastly more BNS detections to significantly greater distances. Given current rate estimates and the strong preference for BNS systems in the initial mass function, BNS mergers could dominate CBC detections in the xG era \citep[cf.][]{2024ApJ...969..108B}. The expected large number of BNS detections, combined with their relative insensitivity to redshift evolution (unlike BBH/NSBH systems), makes BNS mergers strong candidates for spectral siren cosmology.

In this paper, we address two key questions:
\begin{enumerate}
    \item \edit1{Is it necessary to account for the redshift dependence of the BNS mass spectrum when addressing the Hubble tension?}
    \item \edit1{At what redshift is the Hubble parameter best measured using the spectral-siren method?}
\end{enumerate}
To explore these questions, we generated multiple astrophysically motivated BNS catalogs using the population synthesis code \texttt{COMPAS} \citep{Stevenson2017, Vigna-Gomez2018, 2019MNRAS.490.5228B, 2022COMPAS}, varying the physical assumptions and initial parameters. \edit1{To address bias from the redshift evolution of the mass spectrum, we can model the fully correlated population of masses and redshifts, denoted as $\pi (M_c, q, z)$. Here, $M_c$ is the source-frame chirp mass, $q$ is the mass ratio, and we exclude spins since we are focusing on binary neutron star mergers, where spin magnitudes are expected to be small (see section 4.3 of \cite{Chattopadhyay_2020}). Low spin magnitudes for BNSs are consistent with galactic observations, see \cite{Zhu2018NSspin} and \cite{Zhu2020NSspin}. However, there is no universally accepted astrophysics-driven model for $\pi (M_c, q, z)$ due to significant uncertainties in the astrophysical processes and quantities that govern the population. Without a model for redshift evolution, analyses often default to a simplifying assumption, treating mass models as redshift-independent, where $\pi (M_c, q, z) = \pi (M_c, q) \pi (z)$, where $\pi (M_c, q)$ and $\pi (z)$ represent the uncorrelated population distributions of $M_c$ and $q$, and $z$, respectively \citep{Farr_2019, PhysRevD.104.062009}. We test this for each BNS catalog:} we simulate observations with xG observatories and estimate cosmological parameters using models of the mass function that \edit1{are uncorrelated} with redshift to assess whether the true values are accurately recovered. \edit1{In this paper, when we describe two or more parameters as uncorrelated, we mean that they are independent.}

We want to clarify that in this paper we do not infer population together with cosmology. Our focus is different: we test whether cosmological parameters inferred from the injected population remain consistent with the assumption of a non-evolving mass distribution. The BNS mass distribution varies significantly with binary physics (see, for example, Fig.~1 in \cite{vanson2024justwindsmodelsbinary}), and there is currently no astrophysically motivated parametric model that covers all BNS mass distributions produced by population synthesis frameworks. Additionally, mass measurements from both electromagnetic and gravitational wave observations are still too sparse to confirm which synthesized mass distribution is correct (see Appendix \ref{sec:comp}). Here, we investigate whether, despite variations in the BNS mass distribution under different initial conditions and binary physics choices in \texttt{COMPAS}, it remains sufficiently redshift-independent to address the Hubble tension.  Ultimately, \edit1{fitting a population model, whether it depends on redshift or not, will increase statistical uncertainty for a fixed number of observations. This reduces the impact of any bias caused by ignoring the redshift evolution of the mass function. Therefore, our estimates are conservative regarding the effect of bias.}

%%%%%%%%%%%%%%%%%%%%%%%%%%%%%%%%%%%%%%%%%%%%%
\section{Redshift Evolution of Features in Mass Distribution}\label{sec:Corr}

There are several key astrophysical clues that suggest that the mass distribution of BNS mergers will evolve much less with redshift than the BBH and NSBH mass distributions. 

Firstly, there are fewer formation channels that contribute to the population of BNS mergers. 
While BBH mergers are anticipated to form through a wide range of channels including dynamical interactions, homogeneous evolution, and classical isolated binary evolution \citep[e.g.,][]{MandelFarmer2022}, BNSs are expected to form primarily (if not exclusively) from isolated binary channels involving at least one common envelope event (see e.g., Fig.~1 from \citealt{Wagg_2022} and, Fig.~6 from \citealt{Iorio_2023}). % also fig. x from van son in prep
The reason behind this is that BNS mergers are not expected to form through the `stable mass transfer channel' due to limitations imposed by the critical mass ratio \edit1{coming from the requirement of mass transfer stability in} this channel \citep[for details, see][]{van_Son_2022,2024A&A...681A..31P}. 
This implies that we primarily need to focus on the redshift dependence and delay-time distribution of the `common envelope channel,' which is expected to produce relatively short delay times peaking around 1 Gyr \citep[e.g.,][]{Fishbach_2023} leading to a BNS merger rate that closely tracks the overall star formation rate. \edit1{In this paper, we vary the common envelope efficiency parameter, $\alpha$, to values of $0.25, ~0.5, ~0.75$, and $1.0$, to account for the uncertainty in the mass and final orbital separation within the energy formalism of the common envelope channel.}

Second, BNS formation (or the yield \edit2{=(the number of BBHs/BNS formed per unit star formation mass}) has been shown to be much less sensitive to, or even unaffected by the formation metallicity. This is in stark contrast to BBH formation, where \edit1{yield} has repeatedly been shown to drop at high metallicities.
This behavior is consistently observed across various population synthesis models with a very broad variation of physics assumptions \citep[including \texttt{BPASS}, \texttt{MOBSE}, \texttt{StarTrack}, \texttt{COMPAS},  \texttt{SEVN}][]{2016MNRAS.462.3302E, 2018MNRAS.480.2011G, 2018MNRAS.474.2959G, 2018A&A...619A..77K, 2019MNRAS.490.3740N, 2021MNRAS.502.4877S, 2022MNRAS.516.5737B, Iorio_2023,Fishbach_2023}. 
Double compact object formation through the common envelope channel is influenced by metallicity-dependent winds, which reduce the progenitor \edit1{Helium} core masses \edit1{\citep{1998A&A...335.1003H, 2000MNRAS.315..543H, 2017A&A...607L...8V, 2020MNRAS.499..873S, 2023A&A...670A..83S}}. For BBHs, lower-mass BHs receive larger natal kicks because these kicks are assumed to scale inversely with \edit1{Helium core} mass \edit1{\citep{zams4}}. Hence this leads to a loss of potential BBHs at high metallicities. 

\edit2{It has long been established that larger natal kicks reduce the yield of BNS systems \citep[e.g.,][]{Giacobbo_2018, 2022MNRAS.516.5737B}. This is because most binaries do not survive the first supernova natal kick, which in most cases unbinds the system \citep{2019A&A...624A..66R}.}

\edit2{It remains unclear whether neutron star (NS) kicks are really correlated with the progenitor mass, but both theoretical \citep{Coleman_2022}, and observational \citep{2023MNRAS.521.2504O} studies don't find any correlation between progenitor or system mass and the NS kick velocity. As noted in \cite{vanson2024justwindsmodelsbinary}, because of this, population studies typically assume that all NS receive natal kicks (often drawn from a Maxwellian distribution with a dispersion of $\sigma = 265~\text{km/s}$, motivated by pulsar proper motion measurements \citep{2005MNRAS.360..974H}). Under this assumption, the BNS merger rate is not dependent on metallicity.}

\edit2{On the other hand, there are also indications that the explosion energy increases with ejecta mass \citep[e.g.,][]{2021Natur.589...29B}. If the ejecta masses are assumed to be correlated with the progenitor mass, and the explosion energy is equated to the kick velocity, this would propose an opposite trend: more massive progenitors lead to larger kicks, and thus we should see an increase of BNS mergers with metallicity.}

\edit2{In the end, supernova natal kicks remain an active area of research, but if the NS kick magnitude turns out to be truly dominated by chaotic processes (i.e. turbulent flows), then the independence of NS kicks on progenitor masses might hold, and we expect the BNS mass distribution will be less impacted by the changing metallicity distribution of star formation across redshifts \citep{2024AnP...53600170C}.}

Given the single formation channel with a relatively short delay time, and an approximately metallicity-independent yield when marginalized over masses, we expect features in the BNS mass spectrum to be far less dependent on redshift compared to those in the NSBH and BBH mass spectra.
As a result, inferring redshift from detector-frame masses under the assumption of a non-evolving mass model becomes less prone to bias.
    
\edit1{However, our logic does not account for an \emph{uncorrelated} BNS mass-redshift population. Therefore, within the population synthesis framework, we vary uncertain astrophysical parameters that are expected to have a potentially significant impact on the population of BNS mergers and may cause the BNS mass distribution to correlate with redshift. We first consider the fiducial model in \texttt{COMPAS}, as outlined in the \texttt{COMPAS} methods paper \citep{2022COMPAS}. The fiducial model assumes Common-envelope efficiency (CE $\alpha$)=1, mass transfer stability parameter for Hertzsprung gap donors ($\zeta_{\text{HG}}$)=6.5, and  width of the Maxwellian distribution of the kick velocity for electron capture supernovae ($\sigma_{\text{ECSN}}$)=30 (km/s). It adopts the `Delayed' prescriptions from \cite{zams4} for the remnant masses and natal kicks, and has an accretion efficiency $\beta$, that is limited by the thermal timescale of the donor star.} The \edit1{other} population synthesis variations considered are:
\begin{enumerate}[nolistsep]
    \item CE $\alpha = 0.25, ~0.5, ~0.75$. 
    \item Prescription for the distribution of remnant-masses, and the accompanying natal kick: \cite{Fryer_2022}, \cite{Mandel_2020}.
    \item $\zeta_{\text{HG}} = 5, ~5.5, ~6$ \citep[see][for more details]{2022ApJS..258...34R}.
    \item Mass transfer efficiency; i.e., the fraction of donated mass that is accreted during stable mass transfer $\beta = 0.25, ~0.5, ~0.75, ~1$.
    \item $\sigma_{\text{ECSN}} = 10, ~200$ (km/s).
\end{enumerate}

To calculate the star formation rate density (SFRD) as a function of redshift and metallicity, we use the cosmic integrator from the \texttt{COMPAS} suite \citep{2019MNRAS.490.3740N}, following the parameters outlined in \cite{2023ApJ...948..105V}.

\edit1{The mass-redshift contour plots for the fiducial model and the \cite{Mandel_2020} variations are shown in Fig.~\ref{Fig:GMM3D}. For other variations, the plots are provided \href{https://github.com/SoumendraRoy/RedevolBNS/tree/main/Make_Plots/Extra_Plots}{here}.} We observe that the correlations between mass and redshift across these models are weak, supporting our assumption that the mass distribution is nearly independent of redshift. {Among all the variations, the \cite{Mandel_2020} mass distribution exhibits the strongest variation with redshift.}

\subsection{Population Synthesis Variations Not Considered that Could Potentially Show Mass-Redshift Correlation}\label{subsec:whenzdep}
\edit1{While our best guess is that the BNS mass function evolves significantly less with redshift compared to the BBH and NSBH functions, the redshift independence of the BNS mass function could be a byproduct of models failing to capture redshift-dependent evolutionary properties. Below, we outline two key processes that could most plausibly introduce a mass-redshift correlation.}
\begin{enumerate}
    \item \edit1{The physics of the common envelope can intrinsically vary with metallicity. We generated mock catalogs using a fixed common envelope efficiency within the framework of the $\alpha-\lambda$ formalism \citep{1976IAUS...73...35V, 1984ApJ...277..355W, 1988ApJ...329..764L, 1993PASP..105.1373I}. However, this formalism might be too simple to adequately capture the variation in mass distribution with redshift \citep{Ivanova_2013}. For instance, BNSs can accrete gas from the circumbinary disk in the post-common-envelope phase, leading to changes in their masses and orbital separations as a function of metallicity and, consequently, redshift.}

    \item \edit1{Recent observations suggest a bimodal kick distribution and natal kicks scaled with observed total mass in high- and low-mass X-ray binary populations \citep{Zhao_2023}. This could potentially make the BNS yield metallicity-dependent (see \cite{vanson2024justwindsmodelsbinary} for details) and, consequently, the BNS mass distribution redshift-dependent.}
\end{enumerate}
\edit1{These physical processes, or others not discussed here, could cause variations in the BNS mass function across redshift within the detection horizon. Such variations might be significant enough to introduce biases in cosmological parameter recovery, even after accounting for parameter estimation uncertainties. Future studies could explore these possibilities in greater detail. Additionally, we leave the future work for development and application of techniques to mitigate these biases, potentially inspired by those proposed by \citet{Golomb2024} for addressing features in the binary black hole mass spectrum.}

%%%%%%%%%%%%%%%%%%%%%%%%%%%%%%%%%%%%%%%%%%%%%%%%%%%%%%%%
\section{Spectral Siren Cosmology with an Evolving BNS Mass Spectrum}\label{Sec:Result}

\edit2{Fig.~\ref{Fig:GMM3D} shows that the chirp mass and mass ratio distribution along the constant redshift line remains approximately the same. The Mandel-M\"uller population exhibits the highest mass redshift correlation as both chirp mass and mass ratio distributions are bimodal there, with different rate densities to show the highest variation with redshift. However to investigate explicitly} whether assuming a redshift-independent BNS mass population biases our cosmological inference; we conduct an injection-recovery campaign by generating mock GW data for BNS mergers. The mock GW data is generated in Cosmic Explorer sensitivity \citep{evans2021cehorizon}, with the injected population assumed to be fiducial and Mandel-M\"uller. We select the Mandel-Müller model because it exhibits the highest mass-redshift correlation (see \edit1{Fig.~\ref{Fig:GMM3D})}. If this model still allows for unbiased cosmological parameter inference, then the other models should also be safe.

    \begin{figure*}[ht]
        \centering
        \includegraphics[height=9cm, width=15.3cm]{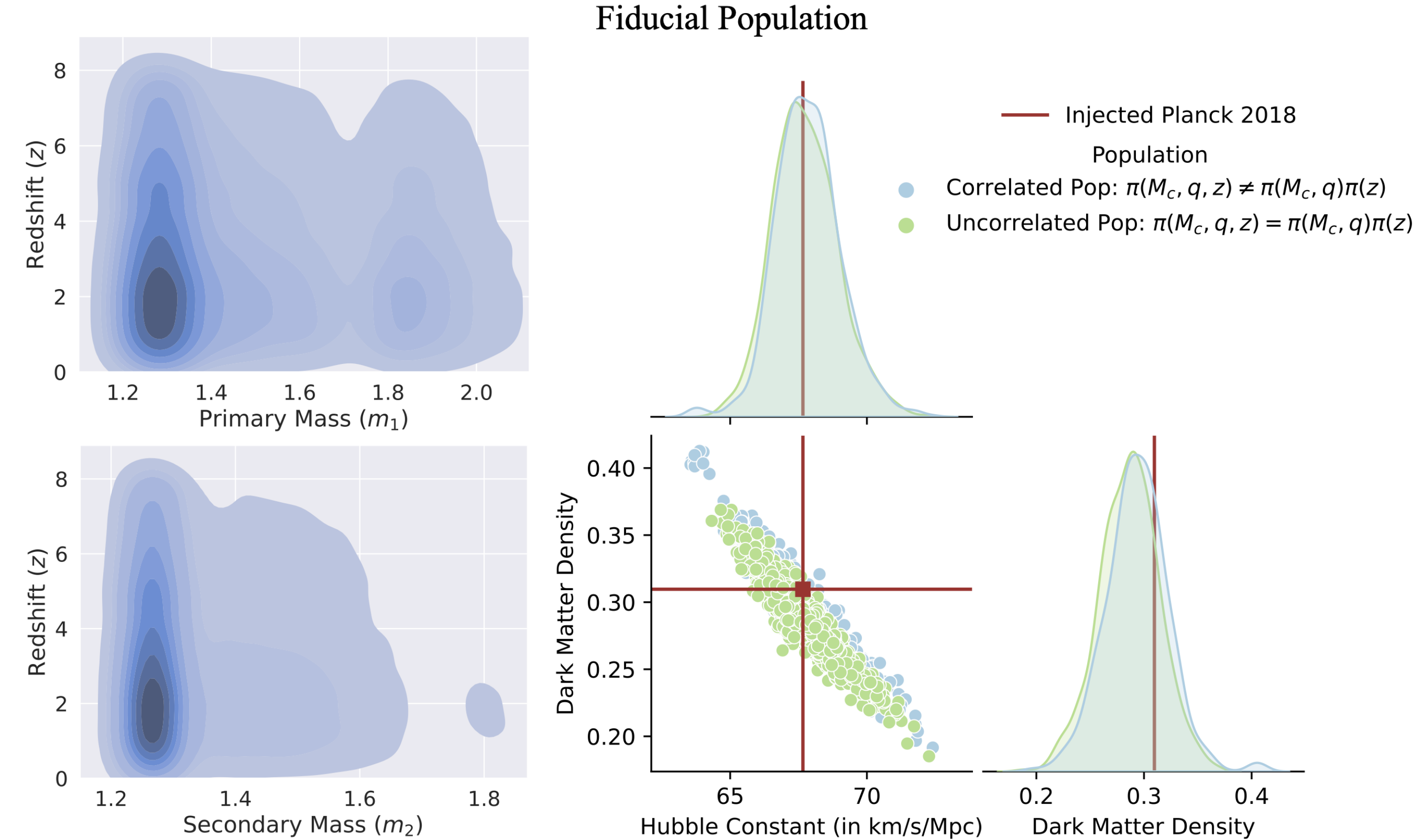}\\
        \vspace{0.5cm}
        \includegraphics[height=9cm, width=15.3cm]{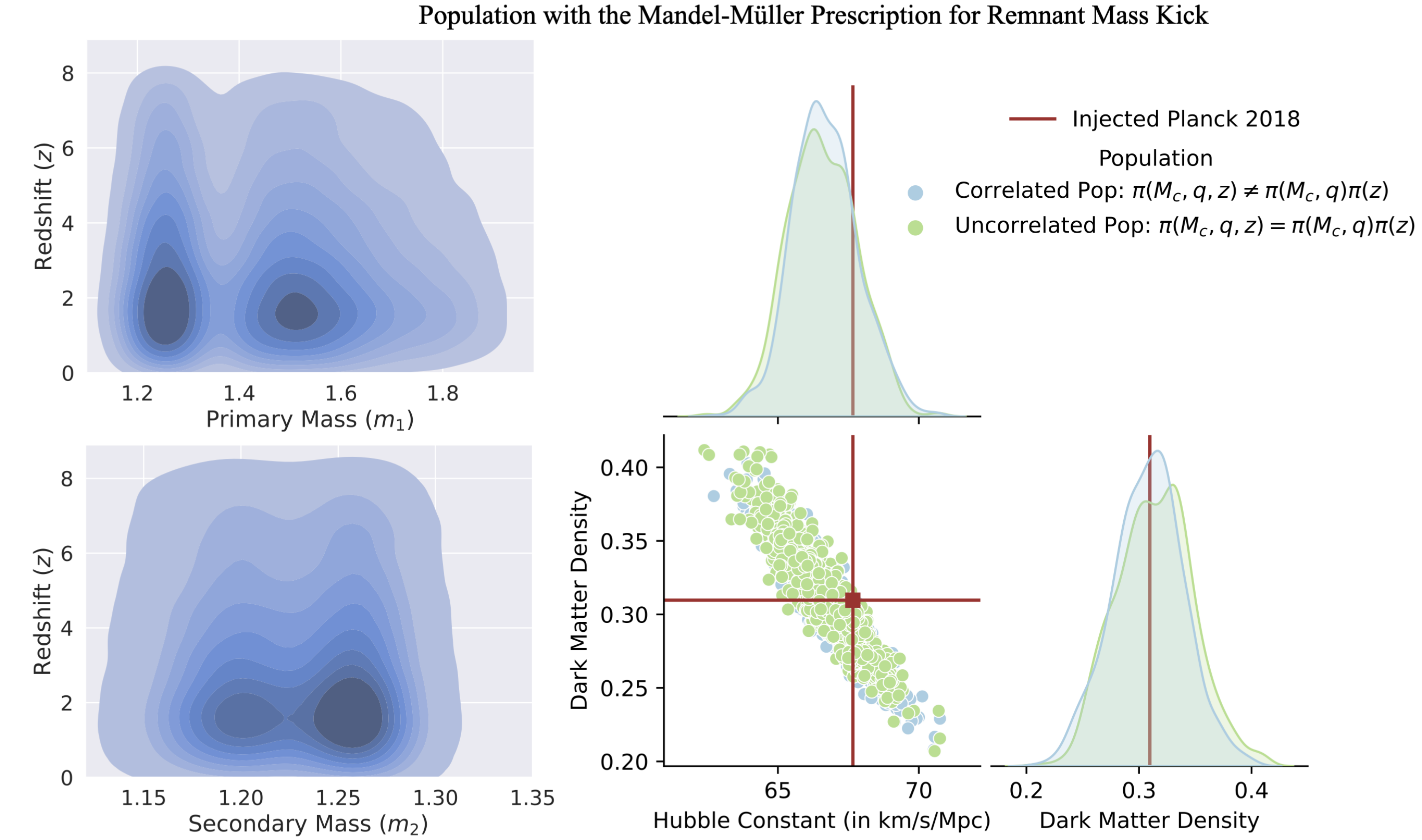}
        \caption{Injection-recovery results for the fiducial model (top) and the \cite{Mandel_2020} remnant-mass and natal kick prescription (bottom). \underline{Left}: Kernel density estimates of the $m_1$ - $z$ and $m_2$ - $z$ populations from \texttt{COMPAS} simulations. In both models, peaks and dips remain stationary with redshift, changing only in absolute merger rate based on star formation, suggesting that the BNS mass function is roughly redshift-independent. \underline{Right}: Inferred matter density and Hubble constant values for both correlated (blue) and uncorrelated (green) populations. Our framework accurately recovers cosmological parameters using both evolving and non-evolving mass models, based on mock data from $800$ binary neutron stars (about 6 days of xG observations at the median BNS merger rate \citep{gwtc-3pop}). For both the fiducial and \cite{Mandel_2020} models, we achieve $\sim 2\%$ constraints on the Hubble constant, \edit1{comparable to \cite{Planck18} and \cite{Riess_2022}, and therefore} sufficient to address the Hubble tension. Code to generate these plots is available \href{https://github.com/SoumendraRoy/RedevolBNS/blob/main/Make_Plots/Pop_Plots.ipynb}{here} and \href{https://github.com/SoumendraRoy/RedevolBNS/blob/main/Make_Plots/Cosmo_Plots.ipynb}{here}.}
        \label{Fig:InjRec}
    \end{figure*}
    
    \begin{figure*}[ht]
        \centering
        \includegraphics[width=8.9cm]{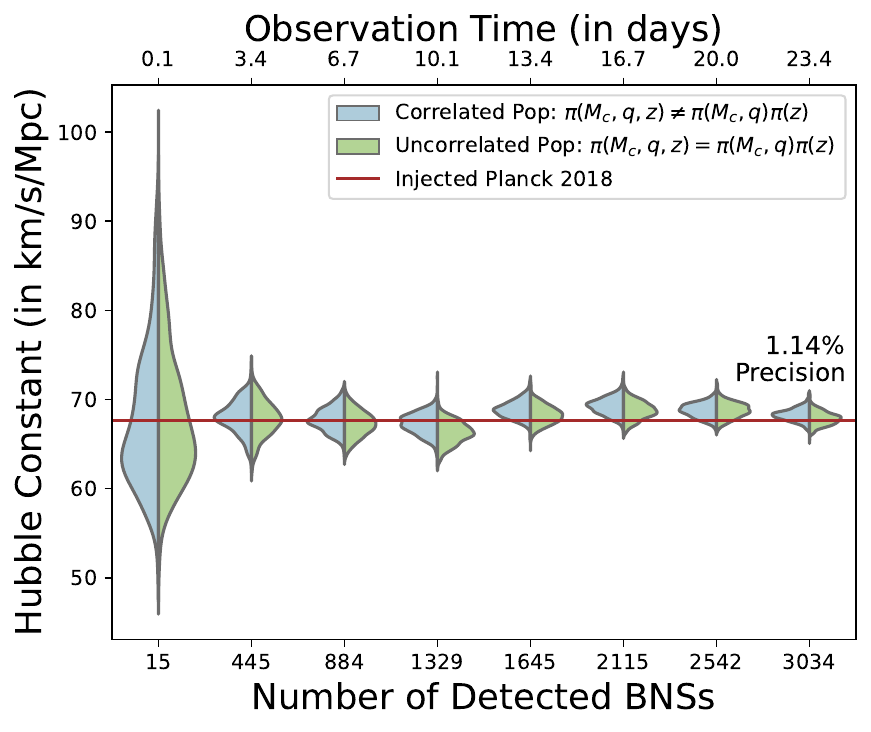}
        \includegraphics[height=7.1cm, width=8.9cm]{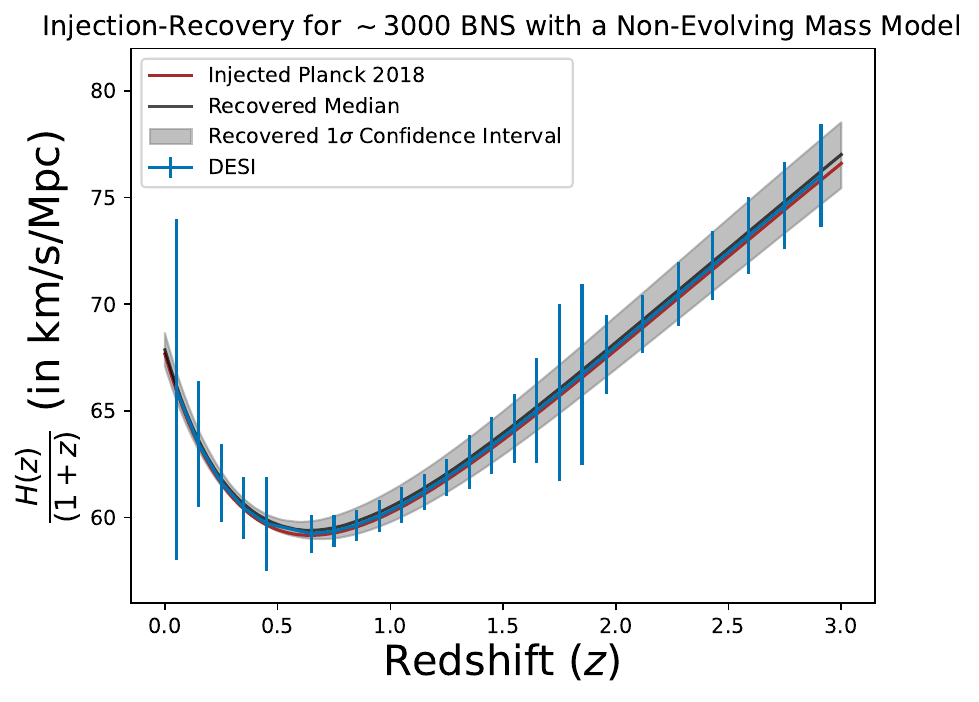}
        \caption{The posterior distribution of the Hubble constant for different numbers of detected BNSs. \underline{Left}: The evolving (blue) and non-evolving (green) mass models are consistent with each other up to approximately $3,000$ BNS observations, which corresponds to one month of data. We achieve percent-level precision in the Hubble constant within this timeframe when we fix the population. For larger numbers of BNS observations, the width of the posterior does not change significantly, indicating it is suppressed by $1/\sqrt{N}$, where $N$ is the number of BNS observations. \underline{Right}: The inferred redshift evolution of the Hubble parameter, $H(z)$, divided by $(1+z)$. $H(z)$ is best constrained near $z=0.4$, achieving less than $0.5\%$ error bar. We accurately recover the true curve when we fix the population to the non-evolving mass model. The forecasted error bars from the DESI \cite{desicollaboration2016desiexperimentisciencetargeting} are also overlaid. Code to generate these plots is available: \href{https://github.com/SoumendraRoy/RedevolBNS/blob/main/Make_Plots/Cosmo_Plots.ipynb}{here}.}
        \label{Fig:H0andHz}
    \end{figure*}

Our population synthesis simulations provide discrete three-dimensional samples of masses and redshifts. We first fit the population with a three-dimensional Gaussian Mixture Model in chirp mass, mass ratio, and redshift space, and then resample from it. We apply a signal-to-noise ratio (SNR) cut of 8 and adopt the Planck 2018 cosmology \citep{Planck18} to convert this into the detected population. Additionally, we incorporate parameter estimation (PE) uncertainties from GWs to obtain samples of detector-frame masses and luminosity distances for each BNS binary. \edit1{See Appendix \ref{sec:MockData} for details.}

We validate our cosmological inference framework using the injected fully correlated population of masses and redshift, represented by $\pi (M_c, q, z)$. Subsequently, we test the assumption of an uncorrelated population, where $\pi(M_c, q, z) \approx \pi(M_c, q)\pi(z)$. \edit1{To model the uncorrelated mass and redshift populations, $\pi(M_c, q)$ \& $\pi(z)$ respectively, we fit them individually using Gaussian Mixture Models. For fitting, we use marginalized samples obtained from the injected distribution: $M_c$ and $q$ marginalized over $z$, and $z$ marginalized over $M_c$ and $q$, respectively}. Here and throughout our models are assuming that the population distribution is \emph{perfectly} known.  This produces maximally precise cosmological inferences and is therefore more sensitive to biases from unmodeled mass-redshift correlation. A realistic analysis that simultaneously fits the population distribution \edit1{(i.e. the moments of population distribution)} \emph{and} cosmology \edit1{(Hubble constant, dark matter density)} will be even less sensitive to these biases at the cost of additional uncertainty in the cosmological parameters. For further details regarding the inference framework, please refer to Appendix \ref{sec:Method}.

Fig.~\ref{Fig:InjRec} presents the injection-recovery results for mock data generated under fiducial assumptions and the Mandel-Müller prescription for 800 detected BNS mergers. For both the fiducial and Mandel-Müller data, the joint posterior distributions of the Hubble constant ($H_0$) and matter density parameter ($\Omega_M$), derived using the assumption of non-evolving mass spectra (uncorrelated case), align with the results obtained when considering the injected population (which may exhibit mass-redshift correlations, i.e., the correlated case). Both methods accurately recover the injected Planck 2018 values with $\sim 2\%$ precision in $H_0$, suggesting that \edit1{the assumption of a redshift independent BNS mass distribution is a valid approximation for its use to estimate the Hubble parameter}.
    
In Fig.~\ref{Fig:H0andHz}, we extend the injection-recovery campaign using fiducial mock data to cover up to one month of
observation time with cosmic explorer. No bias is observed when using a \edit1{redshift-uncorrelated} population model for inference, achieving a $1.14\%$ constraint on $H_0$. For $3,000$ BNS mergers in the fiducial universe, the Hubble parameter, $H(z)$, is best constrained at $z=0.4$, with less than $0.5\%$ bias from any \edit1{mass-redshift correlation}. This suggests that we can make sub-percent level conclusions about the cosmic history of our universe if it follows the fiducial setup of \texttt{COMPAS}.

\cite{chen2024cosmography} have achieved a percent-level constraint on $H_0$ using 50,000 BNS mergers in xG detectors, jointly inferring population and cosmology with a non-evolving uniform mass model. In contrast, by fixing the mass function first to the injected model and then to the non-evolving model, we obtain tighter constraints than \cite{chen2024cosmography}. Based on the constraints from \cite{chen2024cosmography}, we may expect the lack of bias from a non-evolving mass model to remain valid over a year of observation time if we simultaneously infer population and cosmology. The degree of additional uncertainty from this simultaneous inference also depends on the complexity of the population model. Nonetheless, our study suggests that accurately modeling the true mass population marginalized over redshift will not introduce biases in cosmological parameters across various astrophysically motivated BNS catalogs, even at percent-level precision in the Hubble constant.

%%%%%%%%%%%%%%%%%%%%%%%%%%%%%%%%%%%%%%%%%%%%%%%%%%%%%%%%

\section{Conclusion and Future Work}\label{sec:Conclusion}
In this paper, we have evaluated BNS populations as spectral sirens, specifically examining the bias introduced by assuming a redshift-uncorrelated population, $\pi (M_c, q) \pi (z)$, compared to the injected population, $\pi (M_c, q, z)$. Our analysis shows that using a \edit1{redshift-independent} mass model allows us to recover the injected Hubble constant and matter density parameter accurately (see Fig.~\ref{Fig:InjRec}), across a range of realistic population synthesis models, including a fairly extreme example of mass-redshift correlation in the Mandel-M\"uller kick prescription. Our results indicate that a redshift-\edit1{correlated} mass model is not required to achieve a $2\%$ constraint on the Hubble constant under reasonable physical assumptions about the evolution of the BNS mass function with redshift. This level of precision in $H_0$ is sufficient to make a significant contribution toward resolving the Hubble tension. Extending our injection-recovery campaign to one month of Cosmic Explorer data, we have found that a redshift-\edit1{correlated} mass model is not required to achieve sub-percent level precision in the Hubble constant provided BNS mergers in our universe are similar to those of the fiducial setup of \texttt{COMPAS}.

In this study, we have focused on variations of binary physics parameters that are both widely regarded as uncertain, and that we expect could significantly affect the metallicity dependence of the BNS mass distribution. However, our analysis is by no means exhaustive \edit2{(For example, we are considering common envelope $\alpha<1$, as we could not reliably do the injection-recovery campaign in those cases.)}.
The BNS mass distribution may still depend on metallicity, and therefore evolve with redshift, if parameters like the neutron star natal kick distribution or common envelope physics vary intrinsically with metallicity \edit1{(see section: \ref{subsec:whenzdep}).} Nonetheless, we anticipate that the BNS mass distribution is far less prone to redshift evolution than the BBH/NSBH mass distribution (see section \ref{sec:Corr}). \edit2{Hence, we conclude that we can use BNS mergers to infer cosmological parameters without bias to address Hubble tension, given the limited scope of the models used in our study (see the list in section: \ref{sec:Corr}.)}

We have performed the \texttt{COMPAS} simulation using the median rate provided by the LVK collaboration \citep{gwtc-3pop}. However, the BNS rate reported by LVK has a wide uncertainty range, which could affect the accuracy of our mapping from observation time to the number of detections. 
In the ongoing fourth LVK observing run, no BNS mergers have been detected yet. Although the GWTC-4-informed BNS rate is not publicly available, depending on the actual rate, verifying the existence of Hubble tension with Cosmic Explorer may require more or less than one month of observation time, assuming the population model is fixed.

%%%%%%%%%%%%%%%%%%%%%%%%%%%%%%%%%%%%%%%%%%%%%%%%%%%%%%%%

\section{Data Release and Software}
The data used in this work is available on Zenodo: \dataset[10.5281/zenodo.14704635]{https://zenodo.org/records/14704635}. The corresponding codes are available here: \url{https://github.com/SoumendraRoy/RedevolBNS}.

This work made use of the following software packages: \texttt{COMPAS} version \texttt{v02.46.01} \citep{Stevenson2017, Vigna-Gomez2018, 2022COMPAS}, \texttt{astropy} \citep{astropy:2013, astropy:2018, astropy:2022}, \texttt{matplotlib} \citep{Hunter:2007}, \texttt{numpy} \citep{numpy}, \texttt{pandas} \citep{mckinney-proc-scipy-2010, pandas_10045529}, \texttt{python} \citep{python}, \texttt{scipy} \citep{2020SciPy-NMeth, scipy_8079889}, \texttt{corner.py} \citep{corner-Foreman-Mackey-2016, corner.py_4592454}, \texttt{Cython} \citep{cython:2011}, \texttt{h5py} \citep{collette_python_hdf5_2014, h5py_7560547}, \texttt{JAX} \citep{jax2018github}, \texttt{numpyro} \citep{phan2019composable, bingham2019pyro}, \texttt{Jupyter} \citep{2007CSE.....9c..21P, kluyver2016jupyter}, and \texttt{seaborn} \citep{Waskom2021}.

Software citation information aggregated using \texttt{\href{https://www.tomwagg.com/software-citation-station/}{The Software Citation Station}} \citep{software-citation-station-paper, software-citation-station-zenodo}.

%%%%%%%%%%%%%%%%%%%%%%%%%%%%%%%%%%%%%%%%%%%%%

\acknowledgments
\input{ack}

%%%%%%%%%%%%%%%%%%%%%%%%%%%%%%%%%%%%%%%%%%%%%

\appendix
\section{Comparison of Astrophysical Observations with the Redshift Marginalized BNS Mass Population in \texttt{COMPAS}}\label{sec:comp}
Although the BNS mass function is roughly redshift-independent in our models, its shape varies significantly depending on binary physics choices and initial conditions (see, for example, Fig.~1 in \edit1{\cite{vanson2024justwindsmodelsbinary}} and the primary-secondary mass corner plot for simulation variations in Section \ref{sec:Corr}: \href{https://github.com/SoumendraRoy/RedevolBNS/tree/main/Make_Plots/Extra_Plots}{link}). Here, we compare the mass function for the fiducial \texttt{COMPAS} setup with observed astrophysical neutron star mass measurements. We note that this fiducial population aligns with the galactic binary pulsar population inferred by \cite{Farrow2019BNSpop} and with binary neutron star merger populations observed in GW events \citep{gwtc-3pop}.

\begin{figure}[ht]
        \centering
        \includegraphics[height=8cm, width=8.5cm]{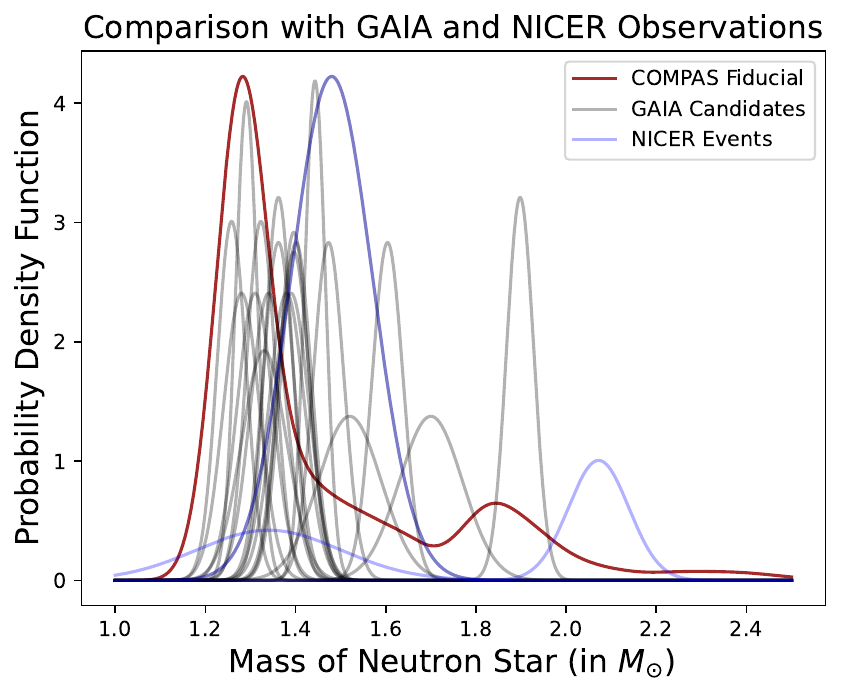}
        \includegraphics[height=8.5cm, width=9.3cm]{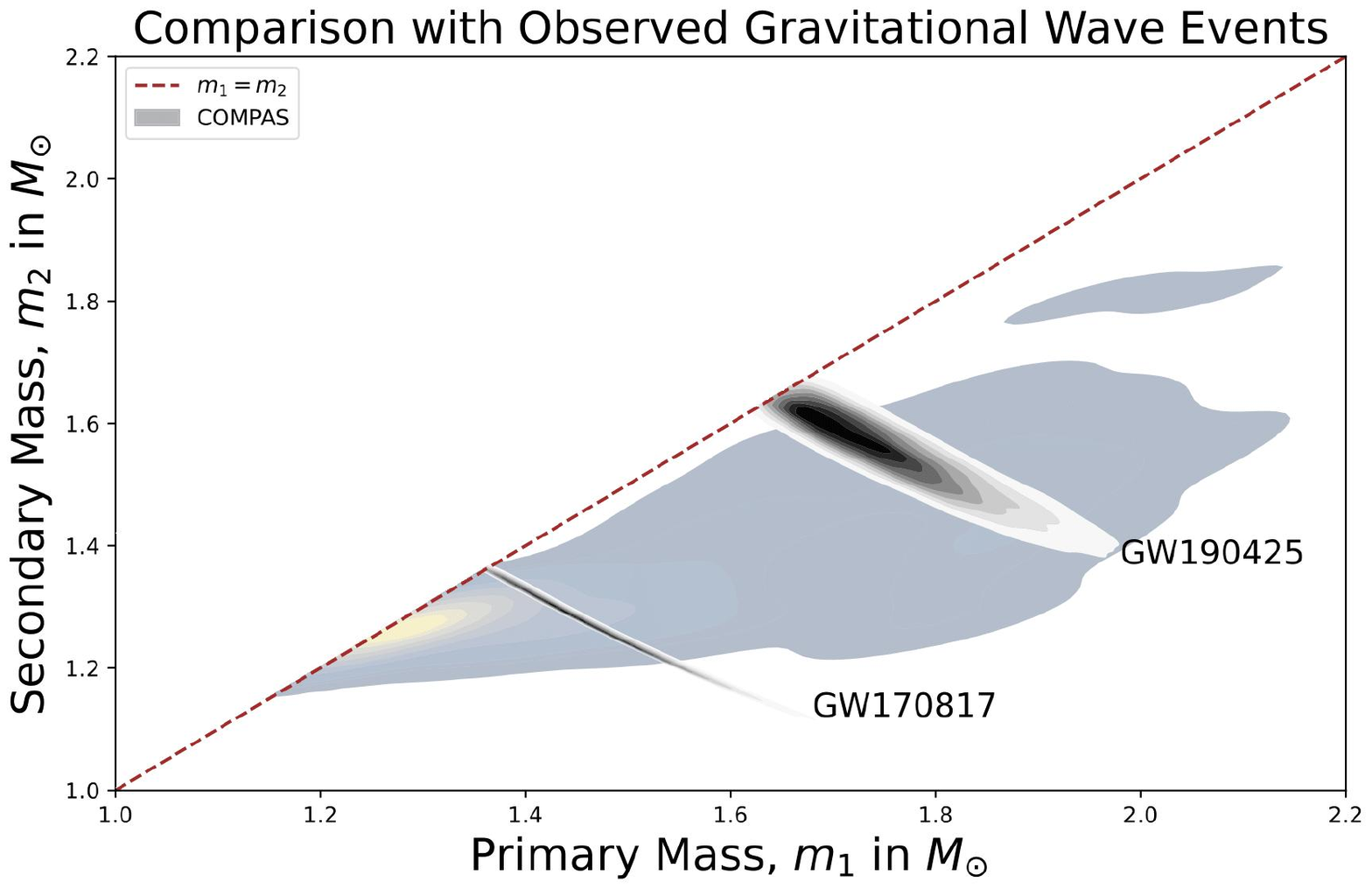}
        \caption{Comparison of individual neutron star mass measurements with the BNS mass population from the fiducial setup in \texttt{COMPAS}. \underline{Left}: Mass measurements from GAIA and NICER are overlaid on the fiducial mass spectrum. Here y-axis is scaled arbitrarily. \underline{Right}: Parameter estimation contours for GW170817 and GW190425 are shown against the fiducial population. Code to generate these plots is available: \href{https://github.com/SoumendraRoy/RedevolBNS/blob/main/Make_Plots/Appendix_Plots.ipynb}{here}.}
        \label{Fig:astro-pop}
\end{figure}

In Fig.~\ref{Fig:astro-pop}, we show neutron star mass measurements for $21$ GAIA candidates \citep{El_Badry_2024} and three NICER observations: PSR J0030+0451 \citep{Riley_2019}, PSR J0740+6620 \citep{Riley_2021}, and PSR J0437–4715 \citep{Choudhury_2024}. We also display parameter estimation contours of the two BNS events detected through GWs so far: GW170817 \citep{Abbott_2017} and GW190425 \citep{Abbott_2020}, analyzed with the \texttt{IMRPhenomPv2\_NRTidal} waveform model and low-spin prior. As shown in Fig.~\ref{Fig:astro-pop}, the range of masses covered by these events are broadly consistent with the fiducial BNS mass spectrum in \texttt{COMPAS}.

%%%%%%%%%%%%%%%%%%%%%%%%%%%%%%%%%%%%%%%%%%%%%

\section{Gaussian Mixture Model Fitting of the True Binary Neutron Star Population}\label{sec:GMMFit}

We draw $10^5$ samples of source-frame chirp mass ($M_c$), logarithm of mass ratio ($\log q$), and redshift ($z$) from \texttt{COMPAS} population using a bounded kernel density estimator with the reflection technique \citep{BDEGaserMuller, BDEHallWehrly}. The bounds are,
\begin{equation*}
    M_c = (-\infty, \infty), ~\log q = (-\infty, 0], ~z = [0, \infty) 
\end{equation*}
While $M_c$ should theoretically be bounded between $[0, \infty)$, in practice, no sample falls below $1M_{\odot}$. Therefore, treating $M_c$ as unbounded is sufficient.

From these, we first split off $50\%$, of the samples data as the ``train'' set, and fit GMMs with varying numbers of components with them. Then we evaluate the likelihood of the rest $50\%$ ``test'' set samples to optimize number of GMM components. We denote the normalized population fitted by the GMM as $f^{\text{GMM}}_{\text{3D}}(M_c, q, z)$. This $f^{\text{GMM}}_{\text{3D}}(M_c, q, z)$ is a representation of the true population generated by \texttt{COMPAS}.

\begin{figure}[ht]
        \centering
        \includegraphics[height=10cm, width=17cm]{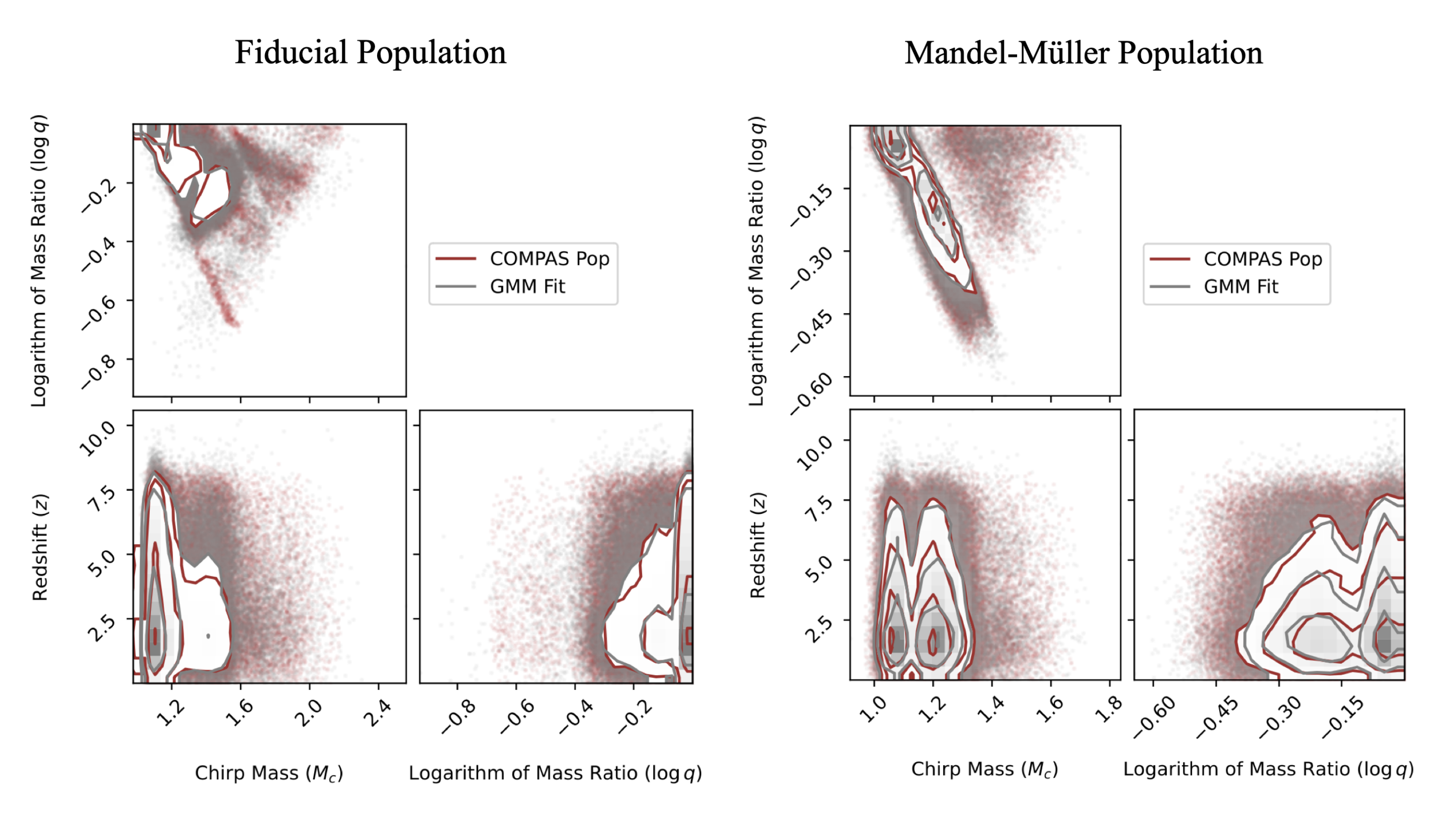}
        \caption{The corner plots show the \texttt{COMPAS} populations of masses and redshifts for both the fiducial setup and the Mandel-M\"uller kick prescription for remnant mass, along with the GMM fit of these populations. In the upper left plot of the fiducial population, fewer than $10$ events following a northwest-to-southeast path ($\log q \approx -0.7$ to $-0.5$) are not captured by the GMM. However, since the chirp mass of these events still falls within the GMM-covered population-and chirp mass is the most precisely measured parameter in cosmological parameter estimation-there is no resulting bias in the cosmological parameters. Code to generate these plots is available: \href{https://github.com/SoumendraRoy/RedevolBNS/blob/main/Make_Plots/Appendix_Plots.ipynb}{here}.}
        \label{Fig:GMM3D}
\end{figure}

\edit2{See Fig.~\ref{Fig:GMM3D} and the \href{https://github.com/SoumendraRoy/RedevolBNS/tree/main/Make_Plots/Extra_Plots}{accompanying material} for the three-dimensional contours and their GMM fits. Animations showing how the mass distribution evolves with redshift can be found \href{https://github.com/SoumendraRoy/RedevolBNS/tree/main/Figures}{here}.}

%%%%%%%%%%%%%%%%%%%%%%%%%%%%%%%%%%%%%%%%%%%%%

\section{Mock Data Generation}\label{sec:MockData}
We summarize the mock data generation process for detector-frame chirp mass ($M_c^{\text{det}}$), mass-ratio ($q$), and luminosity distance ($D_L$) as follows:
\begin{enumerate}
    \item \texttt{COMPAS} provides the number of merging BNSs per unit volume per unit merging time for discrete three-dimensional source-frame samples of masses and redshift. By multiplying the number of merging BNSs per unit volume per unit merging time by $4\pi\frac{dV_c}{dz}\frac{T_{\text{obs}}}{(1+z)}$, we obtain the expected number of BNS mergers, $N_{\text{exp}}$. Here we assume the Planck 2018 cosmology \citep{Planck18}, with $V_c$ and $T_{\text{obs}}$ representing the comoving volume and observation time, respectively. We then perform a Poisson draw with a mean of $N_{\text{exp}}$ to estimate the number of BNS mergers, $N_{\text{obs}}$, that would be observed within the specified observation time, assuming no selection effects,
    \begin{equation*}
        N_{\text{obs}} \sim \PoiDist (N_{\text{exp}})
    \end{equation*}

    \item Given $N_{\text{obs}}$, we generate the true values of $M_c^t$, $q^t$, and $z^t$ by resampling from the true population $f^{\text{GMM}}_{\text{3D}}(M_c^t, q^t, z^t)$:
    \begin{equation*}
        M_c^t, ~q^t, ~z^t \sim f^{\text{GMM}}_{\text{3D}}(M_c^t, q^t, z^t)
    \end{equation*}
    We convert these to the detector-frame values by multiplying $M_c^t$ by $(1+z^t)$ and calculating the luminosity distance, $D_L^t$ for $z^t$ using Planck 2018 cosmology:
    \begin{equation*}
        M^{\text{det},t}_c = M_c^t(1+z), ~D_L^t = D_L^t(z^t; ~\text{Planck 2018})
    \end{equation*}
    \edit1{Here, the Planck 2018 cosmology assumes a flat $\Lambda$CDM model with $H_0=67.66$~km/s/Mpc and $\Omega_M=0.31$ \citep{Planck18}. Other cosmological parameters are not relevant within the redshift range considered.}
    
    We sample the Finn and Chernoff parameter \citep{Finn-n-chirnoff}, $\Theta^t$, from a $\texttt{Beta}(2,4)$ distribution, which approximates the actual distribution closely:
    \begin{equation*}
        \Theta^t \sim \texttt{Beta}(2,4)
    \end{equation*}
    
    \item To obtain PE samples of $M_c^{\text{det},t}$, $q^t$, and $D_L^t$ for each BNS observation, we should create a `true' waveform and inject it into the noise of an xG detector. The resulting signal with added noise provides us `mock' GW data, and we perform a full PE analysis on it. Given the computational expense and the lack of fully established techniques for PE in the next-generation multi-detector setup, we resort to an approximate PE approach \citep{Fishbach_2020, FairhustSimplePE, Farah_2023, essick2023dagnabbitensuringconsistencynoise}.

    We first calculate the true SNR, $\rho^t$, using $M^{\text{det},t}_c$, $q^t$, $\Theta^t$, and $D_L^t$ for each BNS merger:
    \begin{equation}\label{eq:SNR}
        \rho^t = \Bigg[4\int_0^{\infty} df \frac{|h(M^{\text{det},t}_c, q^t, \Theta^t, D_L^t; f)|^2}{S_n(f)} \Bigg]^{\frac{1}{2}}
    \end{equation}
    Here, $h(M^{\text{det},t}_c, q^t, \Theta^t, D_L^t; f)$ is the waveform model, and $S_n(f)$ is the power spectral density of Cosmic Explorer with a $40$ km baseline \citep{Srivastava2022CEpsd}\footnote{For the latest power spectral density data, refer to the URL: \url{https://dcc.cosmicexplorer.org/CE-T2000017/public}}.  We use the single-spin precessing waveform model \texttt{IMRPhenomPv2} \citep{HusaIMRPhenomPv2, KhanIMRPhenomPv2} with all individual spin components set to zero. Evaluating the integral in equation \eqref{eq:SNR} is computationally intensive, especially across the parameter space of BNSs. Therefore, we pre-compute the integral for a two-dimensional grid of face-on masses at a reference distance $D_L^t=1$ Gpc and interpolate for the desired masses. We then find $\rho^t$ at our required Finn and Chernoff parameter and luminosity distance by scaling:
    \begin{equation*}
        \rho^t(M^{\text{det},t}_c, q^t, \Theta^t, D_L^t) = \frac{\Theta^t}{D_L^t} \times \rho^t(M^{\text{det},t}_c, q^t, \Theta^t=1, D_L^t=1~\text{Gpc})
    \end{equation*}
    We determine the observed SNR, $\rho_{\text{obs}}$, by adding Gaussian noise with a standard deviation of $1$ to $\rho^t$:
    \begin{equation}\label{eq:rhodist}
        \rho^{\text{obs}} \sim \NormalDist (\rho^t,1)
    \end{equation}
    Finally, we apply a selection cut by retaining only BNS mergers with $\rho^{\text{obs}}>8$.
    
    \item For BNS mergers above the selection threshold, we aim to create and sample from the joint posterior distribution of the parameters: detector-frame chirp mass, logarithm of mass ratio, Finn and Chernoff parameter, and luminosity distance. The posterior distribution is given by:
    \begin{equation*}
        P(M^{\text{det}}_c, \log q, \Theta, D_L|M^{\text{det},\text{obs}}_c, \log q^{\text{obs}}, \Theta^{\text{obs}}, D^{L,\text{obs}}) \propto \Big|\frac{\partial \rho}{\partial D_L}\Big| \times \pi(M^{\text{det}}_c, \log q, \Theta, \rho) \times
    \end{equation*}
    \vspace{-0.5cm}
    \begin{equation}\label{eq:MDCPos}
        \mathcal{L}(M^{\text{det},\text{obs}}_c, \log q^{\text{obs}}, \Theta^{\text{obs}}, \rho^{\text{obs}}|M^{\text{det}}_c, \log q, \Theta, \rho)
    \end{equation}
    where $\pi(M^{\text{det}}_c, \log q, \Theta, \rho)$ represents the joint prior over the parameters, and $M^{\text{det},\text{obs}}_c, \log q^{\text{obs}}, \Theta^{\text{obs}}, \rho^{\text{obs}}$ are the observed values of these parameters.

    To draw samples of $M^{\text{det}}_c$, $\log q$, $\Theta$, and $D_L$ from Equation \eqref{eq:MDCPos}, we proceed as follows:

    \begin{enumerate}
        \item \textbf{Calculate Observed PE Uncertainty:} We determine the PE uncertainties by scaling the benchmark values from \cite{Vitale3GPEErrorbars} according to the observed SNR ($\rho^{\text{obs}}$). We denote the error bars on $M^{\text{det}}_c, \log q,$ and $\Theta$ as $\sigma_{M^{\text{det}}_c}, \sigma_{\log q},$ and $\sigma_{\Theta}$, respectively.

        \item \textbf{Draw Observed Quantities:} For each event, draw one sample of the observed quantities:
        \begin{equation}\label{eq:observed}
            M^{\text{det}, \text{obs}}_{c} \sim \NormalDist(M^{\text{det}}_c, \sigma_{M^{\text{det}}_c}), ~\log q^{\text{obs}} \sim \NormalDist_{(-\infty, 0]}(\log q, \sigma_{\log q}), ~\Theta^{\text{obs}} \sim \NormalDist_{[0,0.25]}(\Theta, \sigma_{\Theta})
        \end{equation}
        where for $\NormalDist_{[a,b]}(\mu,\sigma)$ is a truncated normal distribution with mean $\mu$ and standard deviation $\sigma$ within the bounds $[a,b]$. 
        
        One notable difference between our mock data generation process and that of \cite{Fishbach_2020} is that while \cite{Fishbach_2020} assumed $M^{\text{det}}_c$ follows a lognormal distribution to avoid negative values of $M^{\text{det},\text{obs}}_c$, we verified that for our values of  $\sigma_{M^{\text{det}}_c}$, $M^{\text{det},\text{obs}}_c$ is never negative.

         \item \textbf{Generate Multiple Samples:} For each observed event, generate $N_{\text{samp}}$ samples (ranging from 1000 to 8000, depending on the number of events):
        \begin{equation}\label{eq:true}
            M^{\text{det}}_c \sim \NormalDist(M^{\text{det},\text{obs}}_{c}, \sigma_{M^{\text{det}}_c}), ~\log q \sim \frac{\NormalDist_{(-\infty, 0]}(\log q^{\text{obs}}, \sigma_{\log q})}{\Phi\big(\frac{-\log q}{\sigma_{\log q}}\big)}, ~\Theta \sim \frac{\NormalDist_{[0,0.25]}(\Theta^{\text{obs}}, \sigma_{\Theta})}{\Big[ \Phi\big(\frac{0.25-\Theta}{\sigma_{\Theta}}\big)-\Phi\big(-\frac{\Theta}{\sigma_{\Theta}}\big) \Big]}, ~\rho \sim \NormalDist(\rho^{\text{obs}}, 1)
        \end{equation}
        $\Phi(.)$ represents the cumulative distribution for normal distribution.

        \item \textbf{Calculate Luminosity Distance:} Using the samples of $M^{\text{det}}_c, \log q, \Theta$, and $\rho$ from \eqref{eq:true}, determine $D_L$ for each event as:
        \begin{equation}
            D_L = \frac{\Theta}{\rho} \times \rho^t(M^{\text{det}}_c, \log q, \Theta=1, D^{\text{fid}}_L=1 \text{Gpc})
        \end{equation}

        \item \textbf{Reweight Samples:} Reweight the samples by:
        \begin{equation}\label{eq:weight}
            w = \Big|\frac{\partial \rho}{\partial D_L}\Big| = \frac{\rho}{D_L}
        \end{equation}
    \end{enumerate}

    Finally, for each of the $N_{obs}$ BNS mergers, we obtain $N_{\text{samp}}$ samples of $M^{\text{det}}_c$, $\log q$, $\Theta$, and $D_L$ which together represent the posterior distribution of PE using a uniform PE prior.
\end{enumerate}

%%%%%%%%%%%%%%%%%%%%%%%%%%%%%%%%%%%%%%%%%%%%%

\section{Method}\label{sec:Method}

In Appendices \ref{sec:GMMFit} and \ref{sec:MockData}, we discussed generating mock data to represent PE samples in Cosmic Explorer for different true population distributions. Next, we estimate the cosmological parameters, $\Vec{\Omega}_C$, using these datasets. To achieve this, we deconvolve the error bars induced by PE \citep{Bovy_2011}. This process yields the rate-marginalized posterior distribution of $\Vec{\Omega}_C$, incorporating a hierarchical prior proportional to the inverse of the rate \citep{Pdet2-essick}.
    
The posterior distribution of $\Vec{\Omega}_C$ is approximately given by \citep{popgw2, popgw1}:

\newcommand\approxprop{\stackrel{\mathclap{\normalfont\mbox{approx}}}{\propto}}

\begin{equation}\label{pos}
    \text{Posterior of} ~\Vec{\Omega}_C ~~~~ \approxprop ~~~~ \pi (\Vec{\Omega}_C) \xi^{-N_{\text{\text{obs}}}} \prod\limits_i^{N_{\text{\text{obs}}}} \sum\limits_{M_c^{\text{det},i}, q^i, D_L^i} \frac{1}{\pi_{\text{PE}}(M_c^{\text{det}}, q, D_L)} . \frac{\frac{dz}{dD_L}}{1+z} . \pi(M_c, q, z) \Bigg|_{M_c=\frac{M_c^{\text{det}}}{1+z(D_L, \Vec{\Omega}_C)}, ~z=z(D_L, \Vec{\Omega}_C)}
\end{equation}

Here, $\pi (\Vec{\Omega}_C)$ represents the hierarchical prior for $\Vec{\Omega}_C$ and $\xi$ is the selection function corresponding to the SNR cut of $8$ in Cosmic Explorer noise. $M_c^{\text{det},i}$, $q^i$, and $D_L^i$ are the samples from the $i$-th event's PE posterior. We also fully marginalize over the Finn and Chernoff parameter, $\Theta$ since its population is independent of masses and redshift and fixed by isotropy.

We calculate the selection function, $\xi$, by generating injections in the detector frame and performing a Monte Carlo summation in the detector frame \citep{Tiwari_2018, Pdet1-Farr},
\begin{equation}\label{sel}
    \xi ~(\Vec{\Omega}_C) \approx \frac{1}{N_{\text{draw}}}\sum\limits_{M_c^{\text{det},\text{inj}}, q^{\text{inj}}, D_L^{\text{inj}}} \frac{1}{p_{\text{draw}}(M_c^{\text{det}}, q, D_L)} . \frac{\frac{dz}{dD_L}}{1+z} . \pi(M_c, q, z) \Bigg|_{M_c=\frac{M_c^{\text{det}}}{1+z(D_L, \Vec{\Omega}_C)}, ~z=z(D_L, \Vec{\Omega}_C)}
\end{equation}
Here, $p_{\text{draw}}$ is the distribution from which the injections are sampled, and $M_c^{\text{det},\text{inj}}, q^{\text{inj}}, D_L^{\text{inj}}$ are the injections that meet the detection threshold of the SNR cut of $8$. In Equations \eqref{pos} and \eqref{sel}, $dz/dD_L$ and $1/(1+z)$ are Jacobians to convert the population density in the source frame into a density over the samples in the detector frame. We also ensure that the effective number of samples is greater than four times the number of observed BNS mergers \citep{Pdet1-Farr, Pdet2-essick}.

For the correlated case, we use the injected population $f^{\text{GMM}}_{\text{3D}}(M_c, q, z)$ as described in Appendix \ref{sec:GMMFit}:
\begin{equation}\label{eq:corrpop}
    \pi_{\text{corr}}(M_c, q, z) = \pi_{\text{inj}}(M_c, q, z) = f^{\text{GMM}}_{\text{3D}}(M_c, q, z)
\end{equation}
where $\pi_{\text{corr}}(M_c, q, z)$ and $\pi_{\text{inj}}(M_c, q, z)$ represent the correlated and injected populations, respectively.

    \begin{figure}[ht]
        \centering
        \includegraphics[height=13cm, width=17cm]{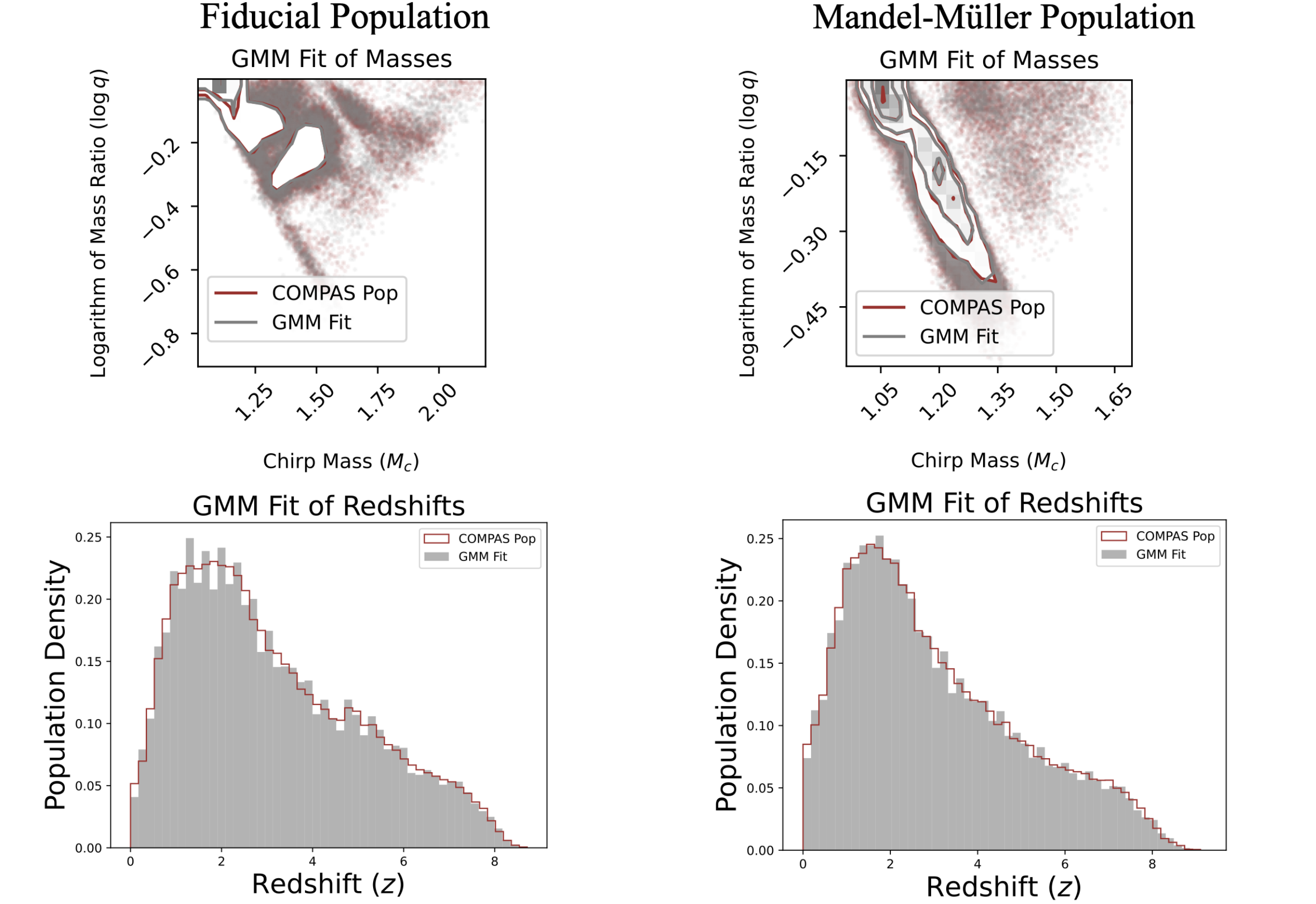}
        \caption{The corner plots and histograms display the marginalized distributions of masses and redshift for both the fiducial and Mandel-Müller populations, along with their GMM fits. Code to generate these plots is available: \href{https://github.com/SoumendraRoy/RedevolBNS/blob/main/Make_Plots/Appendix_Plots.ipynb}{here}.}
        \label{Fig:GMM2p1D}
    \end{figure}
    
For the uncorrelated case in Fig.~\ref{Fig:GMM2p1D}, we model the mass population, marginalized over redshift, using a two-dimensional Gaussian Mixture Model, $f^{\text{GMM}}_{\text{2D}}(M_c, q)$, and the redshift population, marginalized over masses, using a one-dimensional Gaussian Mixture Model, $f^{\text{GMM}}_{\text{1D}}(z)$. The uncorrelated population is then given by:
\begin{equation}\label{eq:uncorrpop}
    \pi_{\text{uncorr}}(M_c, q, z) = \Big(\int_{\forall z}dz~\pi_{\text{inj}}(M_c, q, z)\Big) \Big(\int_{\forall M_c, q}dM_c~dq~\pi_{\text{inj}}(M_c, q, z)\Big) = f^{\text{GMM}}_{\text{2D}}(M_c, q) f^{\text{GMM}}_{\text{1D}}(z)
\end{equation}
Here, $\pi_{\text{uncorr}}(M_c, q, z)$ represents the uncorrelated mass-redshift population. The population drops to zero beyond mass ratio $1$ and below redshift $0$. The GMM underpredicts the probability density near these hard cutoffs. To compensate for this, we use the reflection technique \citep{BDEGaserMuller, BDEHallWehrly}.
    
To model the distance ($D_L$)–redshift ($z$) relationship, we adopt a flat $\Lambda$CDM cosmology with a radiation density parameter set to zero:
\begin{equation}\label{eq:cosmo}
    D_L = \frac{c}{H_0}(1+z) \int_0^z dz^{\prime} \frac{1}{\sqrt{\Omega_M(1+z^{\prime})^3+(1-\Omega_M)}}
\end{equation}
We assume uniform priors for the dimensionless Hubble constant ($H_0/100 ~\text{km s}^{-1} \text{Mpc}^{-1} $), ranging from $0.2$ to $1.5$, and for the matter density parameter ($\Omega_M$), ranging from $0$ to $1$. These priors are applied to both evolving and non-evolving mass models. We use the No-U-Turn Sampler (NUTS) in combination with \texttt{NumPyro} \citep{phan2019composable, bingham2019pyro} and \texttt{JAX} \citep{jax2018github} for sampling the cosmological parameters. \texttt{JAX} enables compatibility with automatic differentiation, facilitating efficient sampling.

\bibliographystyle{aasjournal.bst}
\bibliography{main}

\end{document}

%% file: authors.tex
\newcommand{\sbu}{Department of Physics and Astronomy, Stony Brook University, Stony Brook, NY 11794, USA}
\newcommand{\cca}{Center for Computational Astrophysics, Flatiron Institute, 162 Fifth Avenue, New York, NY 10010, USA}
\newcommand{\nwu}{Center for Interdisciplinary Exploration and Research in Astrophysics, Northwestern University, 1800 Sherman Ave, Evanston, IL 60201}
\newcommand{\pu}{Department of Astrophysical Sciences, Princeton University, 4 Ivy Lane, Princeton, NJ 08544, USA}
\newcommand{\collab}{Collaborators' Universities}

\author[0000-0001-9295-5119]{Soumendra Kishore Roy}
\affiliation{\sbu}

\author[0000-0001-5484-4987]{Lieke A.~C.~van~Son}
\affiliation{\cca}
\affiliation{\pu}

\author[0000-0002-7322-4748]{Anarya Ray}
\affiliation{\nwu}

\author[0000-0003-1540-8562]{Will M. Farr}
\affiliation{\sbu}
\affiliation{\cca}

%% file: ack.tex
The authors thank Jose Mar\'ia Ezquiaga, Max Isi, Claire Lamman, Utkarsh Mali, Ilya Mandel, Suvodip Mukherjee, and Aditya Vijaykumar for valuable discussions and their contributions in shaping this manuscript.

SKR thanks the Center for Computational Astrophysics at the Flatiron Institute for hospitality while this research was carried out. The computations in this work were, in part, run at facilities supported by the Scientific Computing Core at the Flatiron Institute, a division of the Simons Foundation. AR acknowledges support from the National Science Foundation award PHY-2207728.

This material is based upon work supported by NSF’s LIGO Laboratory which is a major facility fully funded by the National Science Foundation. This research has made use of data or software obtained from the Gravitational Wave Open Science Center (gw-openscience.org), a service of LIGO Laboratory, the LIGO Scientific Collaboration, the Virgo Collaboration, and KAGRA. LIGO Laboratory and Advanced LIGO are funded by the United States National Science Foundation (NSF) as well as the Science and Technology Facilities Council (STFC) of the United Kingdom, the Max-Planck-Society (MPS), and the State of Niedersachsen/Germany for support of the construction of Advanced LIGO and construction and operation of the GEO600 detector. Additional support for Advanced LIGO was provided by the Australian Research Council. Virgo is funded, through the European Gravitational Observatory (EGO), by the French Centre National de Recherche Scientifique (CNRS), the Italian Istituto Nazionale di Fisica Nucleare (INFN) and the Dutch Nikhef, with contributions by institutions from Belgium, Germany, Greece, Hungary, Ireland, Japan, Monaco, Poland, Portugal, Spain. The construction and operation of KAGRA are funded by Ministry of Education, Culture, Sports, Science and Technology (MEXT), and Japan Society for the Promotion of Science (JSPS), National Research Foundation (NRF) and Ministry of Science and ICT (MSIT) in Korea, Academia Sinica (AS) and the Ministry of Science and Technology (MoST) in Taiwan. 

This paper carries LIGO document number LIGO-P2400446.